\documentclass[sigconf]{acmart}

\usepackage{amsmath}    
\usepackage{amsfonts}
\usepackage{graphicx}
\usepackage{textcomp}
\usepackage{xcolor}

\usepackage{algorithmicx}           %
\usepackage{algpseudocode}          %
\usepackage{algorithm}              %
\usepackage{comment}                %
\usepackage{enumitem}               %
\usepackage{listings}               %
\usepackage{subcaption}             %
\usepackage{xspace}                 %
\usepackage{tabularx}               %
\usepackage{makecell}               %
\usepackage[raggedrightboxes]{ragged2e} %
\usepackage{svg}                    %
\usepackage{pdfpages}               %
\usepackage[htt]{hyphenat}          %

\definecolor{highlightcolor}{gray}{0.85}
\newcommand{\tool}{Maki\xspace}

\newcommand{\numevaluationprograms}{21\xspace}

\newcommand{\ernstprogramsstudied}{19\xspace}

\newcommand{\nonernstprogramsstudied}{2\xspace}

\newcommand{\bash}{bash\xspace}

\newcommand{\emacs}{emacs\xspace}

\newcommand{\fvwm}{fvwm\xspace}
\newcommand{\gawk}{gawk\xspace}

\newcommand{\gnuplot}{gnuplot\xspace}

\newcommand{\linux}{linux\xspace}
\newcommand{\lua}{lua\xspace}

\newcommand{\xfig}{xfig\xspace}

\definecolor{mGreen}{rgb}{0,0.6,0}
\definecolor{mGray}{rgb}{0.5,0.5,0.5}
\definecolor{mPurple}{rgb}{0.58,0,0.82}
\definecolor{backgroundColour}{rgb}{0.95,0.95,0.92}

\newcommand{\highlightcode}[1]{\colorbox{highlightcolor}{\texttt{#1}}}
\lstdefinestyle{CStyle}{
    numberstyle=\footnotesize,
    basicstyle=\footnotesize,
    breakatwhitespace=false,
    breaklines=true,
    captionpos=b,
    keepspaces=true,
    numbers=left,
    numbersep=1em,
    xleftmargin=2em,
    framexleftmargin=1.5em,
    showspaces=false,
    showstringspaces=false,
    showtabs=false,
    tabsize=2,
    escapechar=|,
    language=C
}

\AtBeginDocument{%
  }

\copyrightyear{2024}
\acmYear{2024}
\setcopyright{rightsretained}
\acmConference[ICSE '24]{2024 IEEE/ACM 46th International Conference on Software Engineering}{April 14--20, 2024}{Lisbon, Portugal}
\acmBooktitle{2024 IEEE/ACM 46th International Conference on Software Engineering (ICSE '24), April 14--20, 2024, Lisbon, Portugal}
\acmDOI{10.1145/3597503.3623323}
\acmISBN{979-8-4007-0217-4/24/04}
\acmPrice{}

\begin{document}
\title{Semantic Analysis of Macro Usage for Portability}

\author{Brent Pappas}
\email{pappasbrent@knights.ucf.edu}
\affiliation{%
  \institution{University of Central Florida}
  \city{Orlando}
  \country{United States}
}

\author{Paul Gazzillo}
\email{paul.gazzillo@ucf.edu}
\affiliation{%
  \institution{University of Central Florida}
  \city{Orlando}
  \country{United States}
}

\renewcommand{\shortauthors}{Pappas and Gazzillo}

\begin{abstract}
  C is an unsafe language.
Researchers have been developing tools to port C to safer languages such as Rust, Checked C, or Go.
Existing tools, however, resort to preprocessing the source file first, then porting the resulting code, leaving barely recognizable code that loses macro abstractions.
To preserve macro usage, porting tools need analyses that understand macro behavior to port to equivalent constructs.
But macro semantics differ from typical functions, precluding simple syntactic transformations to port them.
We introduce the first comprehensive framework for analyzing the portability of macro usage.
We decompose macro behavior into 26 fine-grained properties and implement a program analysis tool, called \tool{}, that identifies them in real-world code with 94\% accuracy.
We apply \tool{} to \numevaluationprograms programs containing a total of 86,199 macro definitions.
We found that real-world macros are much more portable than previously known.  More than a third (37\%) are easy-to-port, and \tool provides hints for porting more complicated macros.
We find, on average, 2x more easy-to-port macros and up to 7x more in the best case compared to prior work.
Guided by Maki's output, we found and hand-ported macros in three real-world programs.  We submitted patches to Linux maintainers that transform eleven macros, nine of which have been accepted.

\end{abstract}

\ccsdesc[300]{Software and its engineering~Semantics}
\ccsdesc[500]{Software and its engineering~Preprocessors}
\ccsdesc[500]{Software and its engineering~Automated static analysis}

\keywords{macros, C, program analysis}

\maketitle

\section{Introduction}
\label{sec:introduction}
C is an unsafe language with millions of lines of critical software infrastructure implemented in it.
Researchers have been developing (semi-)automated tools to port C to safer language, such as c2rust~\cite{c2rust}, 3c~\cite{3c}, and c2go~\cite{c2go}.
Transforming the original, unpreprocessed code is hard, because C programs are written in two languages, C itself and the C preprocessor language, which is used extensively in real-world C programs~\cite{ernst}.
Existing porting tools instead resort to first preprocessing the source file, then porting the preprocessed code, due to long-standing obstacles caused by preprocessor usage~\cite{c2rustissue16,3cissue400,3cissue40,3cissue439}.

The problem with preprocessing first is that the source code is barely recognizable after preprocessing.
The following is an example of function macro usage from the \lua~\cite{lua} source:
\begin{lstlisting}[style=CStyle,frame=none]
#define ALPHABIT      0
#define MASK(B)       (1<<(B))|\label{ln:example:before:mask}|
#define testprop(c,p) (luai_ctype_[(c)+1]&(p))
#define lislalpha(c) testprop(c,MASK(ALPHABIT))|\label{ln:example:before:definition}|
if(|\highlightcode{lislalpha(ls->current)}|)|\label{ln:example:before:invocation}|
\end{lstlisting}
While the highlighted call to \texttt{lislalpha} on line~\ref{ln:example:before:invocation} looks syntactically like a function call, it is a macro invocation, which \emph{expands} the macro, i.e., performs text substitution of the macro's definition on line~\ref{ln:example:before:definition} while also substituting its parameters into the macro body.
After preprocessing, macro definitions are gone and the macro call is reduced to a lengthy series of arithmetic operations on magic constant values:
\begin{lstlisting}[style=CStyle,frame=none,numbers=none]
if(|\highlightcode{(luai\_ctype\_[(ls->current)+1] \& ((1 << (0))))}|)
\end{lstlisting}
This preprocessed source is what tools end up having to port, losing the macro function abstractions from the original source.

To preserve macro usage mixed with C, without having to preprocess it away, porting tools need to be able to understand macro usage before porting to equivalent constructs in the new language.  For instance, some function-like macros, such as \verb'MASK' on line~\ref{ln:example:before:mask} above, behave like C functions once the code is preprocessed and compiled.  Indeed, prominent coding standards even recommend using a C function instead of a macro when it already behaves like a C function~\cite{cert-prefer-functions-over-macros,linuxcodingstyle}.
Such macro usage has a straightforward transformation: create a function of the same name, infer the type of the arguments(s)~\cite{dietrich}, and put the macro body in a return statement.  For instance, \verb'MASK' would become the following, where the highlighted code is the exact contents of the original macro's body:
\begin{lstlisting}[style=CStyle,frame=none,numbers=none]
int MASK(int B) { return |\highlightcode{(1<<(B))}|; }
\end{lstlisting}
But function-like macros, not part of the C language proper, do not always behave like C functions.
Porting tools cannot, in general, apply the simple syntactic substitution used for \verb'MASK' without first identifying whether it is correct to do so.
Macro semantics differs from C~\cite{cppmanual} in calling convention and scoping rules; macros are call-by-name and have dynamic scoping, whereas C functions are call-by-value and have static scoping.
For instance, \verb'PREPEND_LIST', from the \bash source~\cite{bash-5.2-rc1}, resembles a C function, but it assigns a value to an argument, \verb'elist', a side-effect not possible with C's call-by-value calling convention alone:
\begin{lstlisting}[style=CStyle,frame=none,numbers=none]
#define PREPEND_LIST(nlist, |\highlightcode{elist}|) \
  do { nlist->next = elist; |\highlightcode{elist = nlist}|; } while (0)
\end{lstlisting}
A porting tool converting this macro to a C function, for instance, would need to simulate the macro's call-by-name behavior by changing the argument type to a pointer and adding a dereference in the body to match the behavior of the original macro.

The transformations for \texttt{MASK} and \texttt{PREPEND\_LIST} are \emph{interface-equivalent}, i.e., the abstract functional specification of the macro and its C-function equivalent have the same behavior in terms of inputs and outputs.
But in general, macros can freely violate and modify C syntax~\cite{superc,typechef,garrido}.
For instance, there is no interface-equivalent C function for a macro that expands to only a switch statement's case label, as in this example, also from \lua:
\begin{lstlisting}[style=CStyle,frame=none,numbers=none]
#define vmcase(l) case l:
\end{lstlisting}
Porting this macro to C or any typical programming language would require redesigning the function interface and refactoring the code that invokes it, a substantially more complicated transformation than for the interface-equivalent macros \verb'MASK' and \verb'PREPEND_LIST'.
Interface-equivalent macros, in contrast, are \emph{easy-to-port} macros, since they require simpler, largely syntactic transformations.

With an automated program analysis of macro usage, porting tools would be able to determine what transformations preserve behavior.
But prior work on analyzing macro usage for porting is limited to narrow cases.
Mennie and Clarke find and port only parameter-less macros (called object-like macros~\cite{cppmanual}) that are equivalent to C constant variables~\cite{mennie}.
Our evaluation shows that such macros comprise only 19\%, on average, out of all the \numevaluationprograms program we evaluated, leaving behind a substantial amount of easy-to-port, interface-equivalent macros.
The Visual Studio IDE can also convert macros to \texttt{constexpr} variables and functions~\cite{vs-2017-automatic-macro-refactoring}, but this transformation is purely syntactic, suffering similar limitations~\cite{vs-2017-automatic-macro-refactoring-unaddressed-bug} to Mennie and Clarke's work.
This means that developers must manually check that the conversion is correct, since the transformed declaration's definition will at best have the same syntax as the original macro definition, but not necessarily the same behavior.

We introduce the Macro Inspector Framework, the first comprehensive framework for analyzing the portability of macro usage in C programs.
Our framework enables the automated understanding of how macros affect the C program, so that porting efforts can determine the needs for transforming each macro.
The key insight is that our framework takes into account the macro definition and all its invocation sites, comparing the source code both before and after preprocessing to identify how the macro affects the C program.
The challenge is that macros have great freedom to alter the C AST in myriad ways, making it difficult to determine what specific transformation of the macro would preserve behavior.
Therefore, we decompose the changes macros cause into a set of 26 fine-grained properties and design a set of program analyses to discover which properties hold for each macro.
Using our analytical framework, we study the combinations of properties that enable interface-equivalent transformations and which properties need more complicated refactorings before porting, such as the \verb'vmbreak' macro above. %

To evaluate the Macro Inspector Framework, we implement its program analyses in a tool called \tool{}, which automatically identifies the set of properties held by each macro definition.
We use \tool{} to study macro behavior in \numevaluationprograms real-world C programs, including several taken from a classic study of preprocessor usage~\cite{ernst} and \nonernstprogramsstudied more modern programs: the Linux kernel and \lua, with a total of 86,199 macro definitions.
We found that, surprisingly, macros in real-world code are much more portable than previously understood.
More than a third (37\%) are easy-to-port, interface-equivalent macro definitions that have a one-to-one mapping to a C function.
Compared to prior work, we find, on average, 2x more and, counterintuitively, programs with the most complicated macro usage often have even more easy-to-port, interface-equivalent macros compared to prior work, up to 7x more in the best case.
Lastly, we find that syntactic macros, those that respect C syntax, are usually interface-equivalent and consequently, projects with relatively more syntactic macros have more easy-to-port, interface-equivalent macros.
On a statistically significant sample of the benchmark macro definitions, Maki has 94\% accuracy in identifying properties compared to hand-checked ground truth.

To evaluate the utility of the Macro Inspector Framework, we conducted several case studies hand-porting macros guided by Maki's output on two complete programs from our benchmark (m4 and enscript) and two Linux modules.  Macros identified as interface-equivalent took only minutes to port, while non-interface-equivalent macros took substantially longer.  We also submitted patches for a case study of 11 interface-equivalent macros in Linux source and reported on the discussions that led to the acceptance of 9 of them by developers.
This demonstrates that our framework is useful for helping developers move away from macro usage, even in large, mature C codebases with formal review processes.

In this paper, we make the following contributions:
\begin{itemize}
    \item An analytical framework for preprocessor macros that identifies the portability of macro usage (Section~\ref{sec:framework}).
    \item \tool, a Clang plugin and Python library that implements program analyses to detect the framework's properties (Section~\ref{sec:implementation}).
    \item An evaluation of the macro usage and portability in \numevaluationprograms medium-to-large, real-world C programs (Section~\ref{sec:evaluation}).
    \item Case studies of how we used the framework to hand-port macros in several example programs (Section~\ref{sec:case-study}).
\end{itemize}
\tool{}'s complete source code is available online as free and open-source software\footnote{\url{https://github.com/appleseedlab/macro-analyzer}\label{fn:maki-github}} as well as in our publicly-available artifact~\cite{artifact}.

\section{The Macro Inspector Framework}
\label{sec:framework}
\setlength{\abovedisplayskip}{0pt}
\setlength{\belowdisplayskip}{0pt}
\setlength{\abovedisplayshortskip}{0pt}
\setlength{\belowdisplayshortskip}{0pt}

\begin{table*}[htbp]
  \renewcommand{\arraystretch}{1.1}
  \newcommand{\tableindent}[0]{\hspace{2em}}
  \newcommand{\propstyle}[0]{}
  \newcommand{\catstyle}[0]{\emph}
  \fboxsep1pt
    \centering
    \begin{tabular*}{\textwidth}{p{0.25\textwidth}p{0.4\textwidth}p{0.3\textwidth}}
        \textbf{Property} & \textbf{Formal Definition} & \textbf{Description} \\
        \hline
        \hline
        \multicolumn{3}{c}{Interface-Equivalent} \\
        \hline
        \catstyle{Calling-Convention-Adapting} & & \\
        \tableindent
        \propstyle{Modified body} & $\exists e \in \textsc{SideEffectedExprs}(ast), m.ast = e$ & Expands to a side-effected expression \\
        \tableindent
        \propstyle{Modified arguments} & $\exists a \in m.args, \exists e \in \textsc{SideEffectedExprs}(ast), a.ast = e$ & Has a side-effected argument \\
        \tableindent
        \propstyle{Addressed body} & $\exists e \in \textsc{AddressedExprs}(ast), m.ast = e$ & Expands to an addressof (\verb|&|) operand \\
        \tableindent
        \propstyle{Addressed arguments} & $\exists a \in m.args, \exists e \in \textsc{AddressedExprs}(ast), a.ast = e$ & Has an argument that expands to an addressof (\verb|&|) operand \\
        \tableindent
        \propstyle{Unhygienic} & \makecell[l]{$\exists d \in \textsc{LocalDeclRefs}(m.ast), \forall a \in m.args,$ \\ \tableindent $\lnot \textsc{InTree}(d, a.ast), \lnot \textsc{DeclIn}(d, m.ast)$} & Captures a declaration from macro $m$'s caller's environment \\
        \hline
        \catstyle{Scope-Adapting} & & \\
        \tableindent
        \propstyle{Locally defined} & $\lnot  \textsc{DefinedIn}(m,\textsc{GlobalScope}(ast))$ & Defined in a local scope \\
        \tableindent
        \propstyle{Unordered declarations} & $\exists d\in \textsc{DeclRefs}(m.ast), \textsc{DefinedBefore}(m,d)$ & References a declaration defined after macro $m$ \\
        \tableindent
        \propstyle{Unordered expansion type} & $\textsc{DefinedBefore}(m, \textsc{Type}({m.ast}))$ & Expands to an expression whose type is defined after macro $m$ \\
        \tableindent
        \propstyle{Unordered type declarations} & $\exists t\in \textsc{TypeRefs}(m.ast), \textsc{DefinedBefore}(m,t)$ & References a type defined after macro $m$ \\
        \tableindent
        \propstyle{Unordered argument types} & $\exists a  \in m.args, \textsc{DefinedBefore}(m,\textsc{Type}({a.ast}))$ & Has an argument that expands to an expression whose type is defined after macro $m$ \\
        \tableindent
        \propstyle{Unordered macros} & $\exists n\in \textsc{NestedMacros}(m), \textsc{DefinedBefore}(m,n)$ & Invokes a macro defined after macro $m$ \\
        \tableindent
        \propstyle{Condition macro} & $\exists c  \in \textsc{CPPConditionals}, \textsc{InCondition}(m.name, c)$ & Is invoked in a CPP conditional \\
        \tableindent
        \propstyle{Anonymous type} & $\exists e\in \textsc{AnonTypeExprs}(ast), m.ast = e$ & Expands to an expression whose type is unnamed \\
        \tableindent
        \propstyle{Anonymous argument types} & $\exists a  \in m.args, \exists  e\in \textsc{AnonTypeExprs}(ast), a.ast = e$ & Has an argument that expands to an expression whose type is unnamed \\
        \tableindent
        \propstyle{Local argument types} & $\exists a  \in m.args, \exists e\in \textsc{LocalTypeExprs}(ast), a.ast = e$ & Has an argument that expands to an expression whose type is defined in a local scope \\
        \tableindent
        \propstyle{Locally-typed subexpressions} & $\exists e\in \textsc{LocalTypeExprs}(m.ast), e  \in \textsc{SubExprs}(m.ast)$ & Contains a subexpression whose type is defined in a local scope \\
        \tableindent
        \propstyle{Local type} & \makecell[l]{$\exists e\in \textsc{LocalTypeExprs}(ast), m.ast = e,$ \\ \tableindent $\lnot \textsc{DeclIn}(\textsc{Type}(e), m.ast)$} & Expands to an expression whose type is defined in a local scope \\
        \hline
        \hline
        \multicolumn{3}{c}{Non-Interface-Equivalent} \\
        \hline
        \catstyle{Thunkizing} & & \\
        \tableindent
        \propstyle{Void arguments} & $\exists a \in m.args, \textsc{Type}(a.ast) = \texttt{void}$ & Has a void expression argument \\
        \tableindent
        \propstyle{Side-effecting arguments} & \makecell[l]{$\exists a \in m.args, \exists e  \in \textsc{SideEffectExprs}(m.ast),$ \\ \tableindent $\textsc{InTree}(e, a.ast)$} & Has an argument with side-effects \\
        \hline
        \catstyle{Callsite-context-altering} & & \\
        \tableindent
        \propstyle{Unaligned} & $m.ast = null \lor \exists a  \in m.args, a.ast = null$ & Does not align with a single AST node \\
        \tableindent
        \propstyle{Conditional arguments} & $\exists a  \in m.args, \exists e\in \textsc{CondExprs}(m.ast), \textsc{InTree}(a.ast, e)$ & Has an argument that is conditionally evaluated in the body of macro $m$ \\
        \hline
        \catstyle{Nested} & & \\
        \tableindent
        \propstyle{Nested in body} & $\exists n \in \textsc{AllInvocations}, m \in \textsc{NestedMacros}(n)$ & Is invoked in the body of another macro \\
        \tableindent
        \propstyle{Nested in argument} & \makecell[l]{$\exists n \in \textsc{AllInvocations}, \exists a \in n.args,$ \\ \tableindent $m \in \textsc{NestedMacros}(a)$} & Is invoked as an argument to another macro invocation \\
        \hline
        \catstyle{Metaprogramming and Generics} & & \\
        \tableindent
        \propstyle{Control flow} & $m.ast \in \{ \texttt{return}, \texttt{case}, \texttt{continue}, \texttt{break}, \texttt{goto} \}$ & Alters caller's control flow \\
        \tableindent
        \propstyle{Non-expression arguments} & $\exists a  \in m.args, a.ast  \not\in \textsc{Exprs}(ast)$ & Has argument that is not an expression \\
        \tableindent
        \propstyle{Stringizing / Token-pasting} & $\# \in m.tokens \lor \#\# \in m.tokens$ & Uses stringification or token-pasting \\
    \end{tabular*}
    \caption{Macro invocation properties.}
    \label{tab:properties}
\vspace{1em}
\end{table*}

\begin{table*}
    \renewcommand{\arraystretch}{1}
    \centering
    \begin{minipage}[t]{.48\textwidth}
    \begin{tabular}{lp{0.52\linewidth}}
        \textbf{Name} & \textbf{Description} \\
        \hline
        $\textsc{AddressedExprs}(a)$ & Set of expressions in AST $a$ that are the operand of an addressof (\verb|&|) expression. \\
        $\textsc{AllInvocations}$ & Set of all macro invocations. \\ 
        $\textsc{AnonTypeExprs}(a)$ & Set of expressions in AST $a$ whose type is unnamed. \\
        $\textsc{DeclIn}(d, a)$ & Whether declaration $d$ was declared in AST $a$. \\
        $\textsc{DeclRefs}(a)$ & Set of expressions in AST $a$ that are references to declarations. \\
        $\textsc{DefinedBefore}(m, d)$ & Whether macro $m$ is defined before declaration or macro $d$. \\
        $\textsc{DefinedIn}(d, s)$ & Whether declaration or macro $d$ is defined in scope $s$. \\
        $\textsc{CondExprs}(a)$ & Set of short-circuiting expressions (e.g., the ternary operator, or logical and) in AST $a$. \\
        $\textsc{CPPConditionals}$ & Set of CPP static conditionals (e.g., \verb|ifdef|, \verb|defined|, etc.) in program. \\
        $\textsc{GlobalScope}(a)$ & Global scope of AST $a$. \\
    \end{tabular}
    \end{minipage}
    \hspace{0.02\linewidth}
    \begin{minipage}[t]{.48\textwidth}
    \strut\vspace*{-\baselineskip}
        \begin{tabular}{lp{0.52\linewidth}}
        \textbf{Name} & \textbf{Description} \\
        \hline
        $\textsc{InCondition}(s, c)$ & Whether symbol $s$ appears in CPP static conditional $c$. \\
        $\textsc{InTree}(a, b)$ & Whether AST $a$ is a subtree of $b$. \\
        $\textsc{LocalDeclRefs}(a)$ & Set of expressions in AST $a$ that are references to declarations declared in local scopes. \\
        $\textsc{LocalTypeExprs}(a)$ & Set of expressions in AST $a$ whose type is defined at local scope. \\
        $\textsc{NestedMacros}(m)$ & Set of macro invocations or arguments in $m$'s nested invocations. \\
        $\textsc{SideEffectExprs}(a)$ & Set of expressions in AST $a$ with side-effects. \\
        $\textsc{SideEffectedExprs}(a)$ & Set of expressions in AST $a$ that are modified by an assignment expression or the unary increment or decrement operator, or are passed to a function call. \\
        $\textsc{Type}(e)$ & Type of expression $e$. \\
        $\textsc{TypeRefs}(a)$ & Set of expressions in AST $a$ that reference a type declaration. \\
    \end{tabular}
    \end{minipage}
    \caption{Helper functions used in formal definitions of properties.}
    \label{tab:helper-functions}
\end{table*}

\newcommand{\property}[1]{\paragraph*{\textbf{#1}}}

We detail the syntactic and semantic properties of macro usage defined in the Macro Inspector Framework, illustrating them  with examples. %
Table~\ref{tab:properties} lists each property, including its name and formal and informal definitions, that together comprise our framework.
All formal definitions take a macro \textit{m} that contains three fields, tokens of the unpreprocessed macro invocation (\textit{tokens}), the preprocessed macro as an AST if syntactically-valid C (\textit{ast}), and the macro's preprocessed arguments as an AST if syntactically-valid (\textit{args}).

We group properties into \emph{portability categories} that provide guidance to porting tools about what language features are implicated in the macro usage and need to be supported in order to port the macro.
\emph{Calling-convention-adapting} properties concern macro behavior that can be ported by adapting the arguments of the function when ported to a C-like function.   \emph{Scope-adapting} properties involve the use of dynamic scope, and so require modifications to ensure static scope equivalence.
\emph{Definition-adapting} macros are those, like \verb'MASK' (Section~\ref{sec:introduction}), that require only a syntactic change and involve no changes to calling convention or scoping.

Together, macros that are calling-convention-, scope-, or definition-adapting have a one-to-one equivalence with a C-like function and need only minimal changes to port away from preprocessor usage, i.e., they are \emph{interface-equivalent}.  Interface-equivalent macros are relatively easy to port, since they behave like C functions, so identifying macros that are interface-equivalent should help developers more easily port away from macro usage.
In contrast, non-interface-equivalent macros require a redesign of the macro's functional interface, representing more complicated and difficult-to-port macro usage.  We group properties of such macros into \emph{Thunkizing} for those that require converting expressions to functions, i.e., thunks, \emph{Call-site-context-altering} for those that alter the syntax of the macro call-site, \emph{Nested} for nested macro usage, and \emph{Metaprogramming} for macros that perform code generation.

\subsection{Interface-Equivalent Properties}

The \verb'PREPEND_LIST' macro (from Section~\ref{sec:introduction}) causes side-effects on the value of its argument \verb'elist', which is not supported in C or other languages with call-by-value semantics but is with call-by-reference semantics.  We call this behavior the \emph{Modified arguments} property (or \emph{Modified body} when it occurs in the macro body) and provide a formal definition in Table~\ref{tab:properties}.  
A related macro behavior, also not supported by call-by-value, is the use of the C address-of operator (\&).
For example, the \verb|linkgclist| macro below (from \lua) takes the address of its \verb'o' and \verb'p' arguments:
\begin{lstlisting}[style=CStyle]
#define linkgclist(|\highlightcode{o}|,p) \
  linkgclist_(obj2gco(o), |\highlightcode{\&(o)}|->gclist, |\highlightcode{\&(p)}|)
\end{lstlisting}
Argument values and addresses have function-local scope, so the address will not be the same as the parameter passed to the function call with call-by-value semantics.

The \emph{Unhygienic} macro property stems from the same lack of macro scoping that affects address-of.
For example, \verb|ISSET| (from \gawk~\cite{gawk-5.1.1}) expands to an expression that captures the local variable \verb|sp| (highlighted in gray) from its caller's environment, not possible with statically-scoped functions.
\begin{lstlisting}[style=CStyle]
#define	ISSET(opt)	(|\highlightcode{sp}|->fts_options & (opt))
void f() {
  FTS |\highlightcode{*sp}|;
  if (ISSET(FTS_LOGICAL))|\label{line:invoke-isset}|
}
\end{lstlisting}
We define unhygienic macros as those that capture symbols from the function-local scope and use them in the expanded macro invocation site.
When the macro itself is defined inside of the function-local scope it is \emph{locally-defined}.
The formalization checks whether the macro has been defined in any scope but the global scope.

The \emph{unordered declarations, types, type declarations, argument types, and macros} properties all stem from the dynamic scoping of macros.
Rather than requiring symbols to exist at macro define time, as in statically-scoped languages, macros need not check for other symbols being defined until they are invoked.
For example, \verb|open_spline| (line~1), from \xfig~\cite{xfig-3.2.8b} takes a symbol \verb's' that has type \verb'F_spline', which is declared on line~2 after \verb'open_spline':
\begin{lstlisting}[style=CStyle]
#define open_spline(|\highlightcode{s}|) (!(s->type & 0x1))
typedef struct f_spline |\highlightcode{F\_spline}|;
void update_spline(F_spline *spline) {
  if (open_spline(spline)); |\label{line:open_spline-call}|
}
\end{lstlisting}

A \emph{condition macro} is one which is present in a preprocessor static conditional, e.g.., within an \verb|#ifdef|'s condition.
The preprocessor checks if macros are defined, but not functions, so a porting tool handling the condition macro needs to also alter the original program so that all preprocessor conditionals relying on that macro will behave the same after the transformation.

Macros can expand to \emph{anonymous types} and \emph{argument types}.
For example, the macro \verb|TB_FLAGS| defined in \fvwm~\cite{fvwm-2.6.9} expands to an anonymously-typed expression:
\begin{lstlisting}[style=CStyle]
typedef struct { |\highlightcode{struct \{ ... \} flags}|; } TitleButton;
#define TB_FLAGS(tb)              ((tb).flags)
static void SetLayerButtonFlag(..., TitleButton *tb) {
    TB_FLAGS(*tb).has_layer = 1; ... |\label{line:invoke-tb-flags}|
}
\end{lstlisting}
The invocation of \verb|TB_FLAGS| on line~\ref{line:invoke-tb-flags} expands to the expression \verb|((*tb).flags)|, which is described by the anonymous struct type as highlighted in gray.

Related properties exist for invocations that expand to locally-typed expressions (\emph{local type}), accept locally-typed expressions as arguments (\emph{local argument types}), or contain \emph{locally-typed subexpressions}, because macro invocations are untyped and can capture type symbols from the caller's scope.  Functions, however, cannot, because function type signatures only have globally-defined types available in scope.

\subsection{Non-Interface-Equivalent Properties}

Non-interface equivalent macros do not have a one-to-one mapping to C-like function behavior and represent the most complicated uses of macros, requiring refactoring of the original use of the program before mapping them to functions.
For instance, macros may take \emph{void arguments}, i.e., expressions that have no return value, such as a call to a void function.
But void arguments are not legal as parameters to C functions.
Similarly, when an argument to a macro causes a side effect (\emph{side-effecting arguments}), every use of that argument repeats the effect due to call-by-name semantics, whereas in C's call-by-value calling convention, the side effect is only computed once.
For instance, macro \verb|min|, from \emacs~\cite{emacs-28.1}, expands the same argument multiple times:
\begin{lstlisting}[style=CStyle, numbers=none]
#define min(a, b) ((a) < (b) ? (a) : (b))
\end{lstlisting}
If invoked with \verb'min(x++, y++);', the macro would increment either \verb'x' or \verb'y' twice, whereas a syntactically-similar C function would only trigger the increment once, when the arguments get evaluated.

Macro invocations operate at the raw token level and thus do not need to expand to or accept complete syntactic constructs, e.g. statements or expressions, as arguments.
When a macro's expansion does not correspond to a complete syntactic construct, we identify it as \emph{unaligned}.
Take the following macro expansion:
\begin{lstlisting}[style=CStyle]
#define ADD(a, b) a + b
4 * ADD(5, 6) |\label{ln:invoke:add}|
\end{lstlisting}
A function version of \verb'ADD' would evaluate the addition first, leaving the multiplication, \verb'4 * 11', while the macro version results in an evaluation of the multiplication first, due to operator precedence on the expanded expression,  \verb|4 * 5 + 6|, a common pitfall in preprocessor usage~\cite{cppmanual-operator-precedence}

If a macro invocation expands one of its arguments into a short-circuiting expression (e.g., a ternary expression or logical conjunction expression), then it has \emph{conditional arguments} and may never evaluate that argument.
For instance, the macro \verb|AND| expands to \verb|((0) && (*x))| which, due to short-circuiting, would never raise a null pointer fault in spite of the dereference of NULL with \verb'*x':
\begin{lstlisting}[style=CStyle]
#define AND(a, b) ((a) && (b))
int *x = NULL;
AND(0, |\highlightcode{*x}|);|\label{line:and-call}|
\end{lstlisting}
If ported to a function with call-by-value semantics, in contrast, then the \verb'*x' dereference would always raise the fault.

A macro can alter its caller's \emph{control flow} directly by expanding to any of the \verb|continue|, \verb|break|, \verb|goto|, or \verb|return| keywords.
An expansion to a \verb'goto', for instance, pierces the function abstraction by enabling the callee to return to any point in the caller, because the \verb'goto' will be expanded by the macro into the body of its caller.
\emph{Nested macros} are those that are called by other macros.
This may occur in the definition of the macro, for instance \verb'lislalpha' in Section~\ref{sec:introduction}, or in a parameter at the macro's invocation site.
Unlike functions, macros do not need to be passed expressions as arguments and can instead be passed any syntactic construct, such as statements, declarations, or even unaligned constructs.
These \emph{non-expression arguments} are not supported in C functions.
Macro invocations can use the \emph{stringification} and \emph{token-pasting} operators to manipulate their arguments' tokens, which generate string literals and fresh language tokens at macro expansion time.
This kind of code generation allows for metaprogramming, such as reflection to print error messages or system-dependent type names.

\section{Implementation}
\label{sec:implementation}

\begin{figure}
    \centering
    \includegraphics[width=\linewidth]{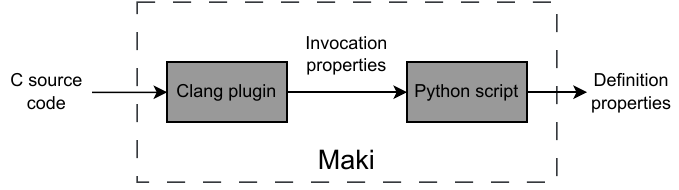}
    \caption{\tool architecture diagram.}
    \label{fig:arch-diagram}
\end{figure}

We have implemented the Macro Inspector Framework in a new tool called \tool which performs automated program analysis to identify macro usage properties.  
\tool consists of a Clang plugin comprised of 2,180 lines of C++ code; and 1,519 lines of Python code.
Figure~\ref{fig:arch-diagram} presents \tool's architecture.
The Clang plugin analyzes macro invocations in individual C source files, so to analyze a complete program we first intercept its build system using Clang's \verb|intercept-build| utility and use the resulting compile commands to analyze all files in the program.
These results can then be passed to \tool's Python scripts to determine which properties apply to each of the program's macro definitions.

The Clang plugin works by hooking into the Clang preprocessor and AST to determine which macros are syntactically aligned.
\tool{} has methods that implement each property's formalism by inspecting the C AST and a trace of preprocessor expansions.
The Python scripts read these results into memory and use this information to determine which properties each macro definition satisfies.
Based on these results, we can determine which portability category applies to each macro definition.
\tool{}'s source code is available both on GitHub ~\footref{fn:maki-github} and in our public artifact~\cite{artifact}.

\section{Evaluation}
\label{sec:evaluation}
\nocite{datawrapper}

\begin{figure*}
\centering
\includegraphics[height=4in]{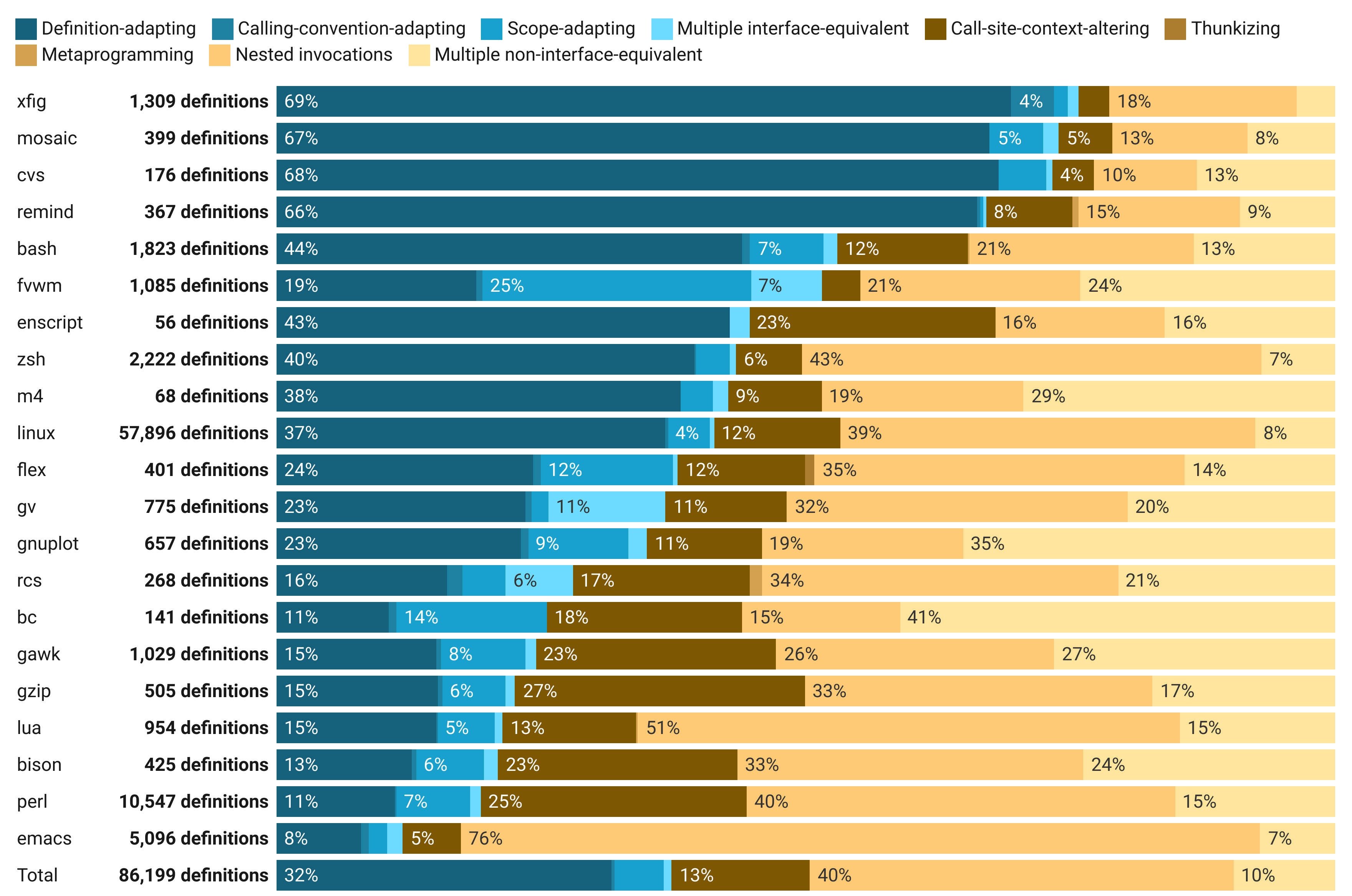}
\caption{Percent of macro definition adaptation categories.}
\label{fig:definition-transformation-types}
\end{figure*}

We use \tool to analyze real-world macro usage portability and answer the following research questions:
\renewcommand{\theenumi}{\textbf{RQ\arabic{enumi}}}
\renewcommand{\labelenumi}{\theenumi}
\begin{enumerate}
    \item (Portability) How much portability is there in real-world macro usage?
    \item (Comparison) How does \tool's ability to find portable macros compare to that of prior work?
    \item (Alignment and Portability) How does syntactic macro usage influence portability?
    \item (Runtime) How quickly does \tool analyze macro usage?
    \item (Accuracy) How accurately does \tool identify properties?
\end{enumerate}
All summary data is in our free, publicly-available artifact~\cite{artifact}.

\subsection{Benchmark Program Selection}
We draw \ernstprogramsstudied programs to analyze from Ernst et al.'s analysis of C preprocessor usage~\cite{ernst}\footnote{bash, bc, bison, cvs, emacs, enscript, flex, fvwm, gawk, gnuplot, gv, gzip, m4, mosaic, perl, rcs, remind, xfig, and zsh} but omit the remaining seven which either contain C++ code\footnote{gcc, ghostscript, and gnuchess} or which we could not build due to missing dependencies\footnote{zephyr, workman, and RasMol}.
To augment the benchmarks, we also add \nonernstprogramsstudied very large, modern programs as well: the Linux kernel and Lua.
Across all programs in our benchmark, there are 86,199 macro definitions.

\subsection{Experimental Setup}
We ran our experiments on a server with 2 AMD EPYC 7742 64-Core hyper-threaded processors, for a total of 256 available processes at one time, with 512GB RAM.
We download and extract each program, and intercept its build system with Clang's \verb|intercept-build| tool to obtain the specific compile commands used to compile each source file.
We pass each of these compile commands to \tool's Clang plugin to determine which properties each macro invocation in the program satisfies.
We run the Clang plugin on eight cores for all programs except for Linux, for which we use 32 cores due to the kernel's size.
Finally, we run \tool's Python scripts to determine which properties apply to each macro definition.
We only examine invocations of macros defined in the programs themselves and not invocations of macros defined in any system header or library files, since they are not part of the application.
We reimplement the macro usage analysis from Mennie and Clarke, which ports only constant, object-like macros, in \tool{}, and measure how often they occur in each program.

\subsection{RQ1: Portability}
\label{sec:rq-2}

We use \tool{} to evaluate each of the macro definitions across all programs in our benchmark suite, recording which properties hold for each macro.
For each macro, we use its properties to assign it to a portability category, as indicated in Table~\ref{tab:properties}.
Figure~\ref{fig:definition-transformation-types} presents our results.
Each program has a segmented bar chart that represents the percentage distribution of macro definitions in each portability category.
When there are multiple invocations of the same definition that have differing portability categories, we record them as having either multiple interface-equivalent properties or multiple non-interface equivalent.
The interface-equivalent portability categories are color-coded by blue shades, whereas non-interface equivalent are yellow shades.
The programs are sorted by the highest percentage of interface-equivalent macro definitions.

Programs have widely-varying macro portability, ranging from 12\% to as much as 76\% interface-equivalent macros.
The largest program with the most (57,896) macro definitions is Linux with 41\% being easily-portable.
On average, over all 86,199 macros, 37\% are easy-to-port, being interface-equivalent.
The program analyses of macro semantics in the Macro Inspector Framework has enabled us to discover that macros in real-world code are much more portable than previously understood.

\subsection{RQ2: Comparison}

\begin{figure}
\centering
\includegraphics[height=4in]{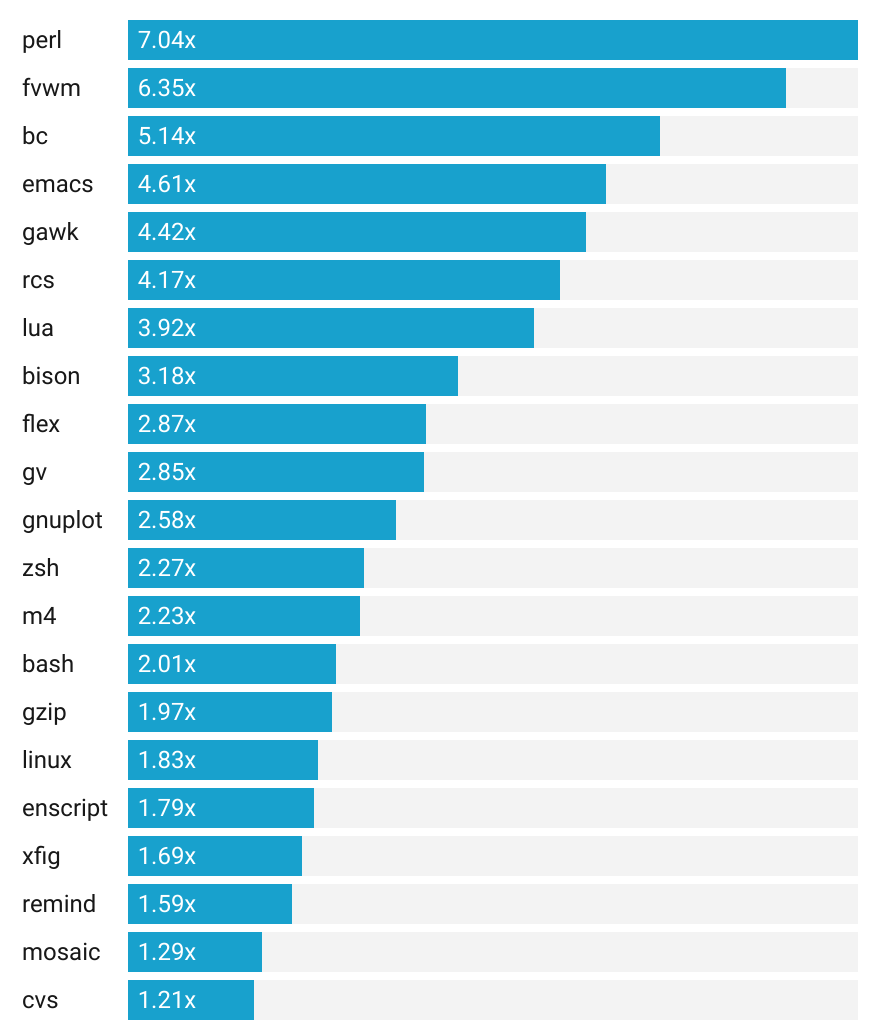}
\caption{How many more easy-to-port, interface-equivalent macro definitions \tool finds over prior work.}
\label{fig:comparision-to-prior-work}
\end{figure}

We measure how many macros are identified as portable according to our reimplementation of Mennie and Clarke's~\cite{mennie} macro usage analysis and compare the number of these macros to the number of interface-equivalent macros identified by our framework's properties implemented in \tool{}.
Figure~\ref{fig:comparision-to-prior-work} presents the relative performance of Mennie and Clarke's tool against \tool, by dividing the number of interface-equivalent macros over the constant, object-like macros identified by Mennie and Clarke.
We sort the results by most to least increase in identified, portable macros.
Prior work transforms an average of 19\%, a minimum of 3\%, and a maximum of 61\% of macro definitions across all the programs we study.
On average, \tool{} finds twice as many portable macros, ranging from 1.21x to 7.04x more across all programs.

Comparing Figure~\ref{fig:definition-transformation-types} to the relative performance results, we see that some of the greatest improvement in portable macro identification occurs in programs with some of the most complex macros usage.
For instance, perl, in which \tool{} identifies 7.04x more portable macros, has some of the most non-interface-equivalent macro definitions.
This indicates that complex macro usage obscures macro portability, and \tool{}'s analyses help tease out the semantic aspects of macro usage in order to identify easily-portable macros.

\subsection{RQ3: Alignment and Portability}

\begin{figure*}
\centering
\includegraphics[width=\textwidth]{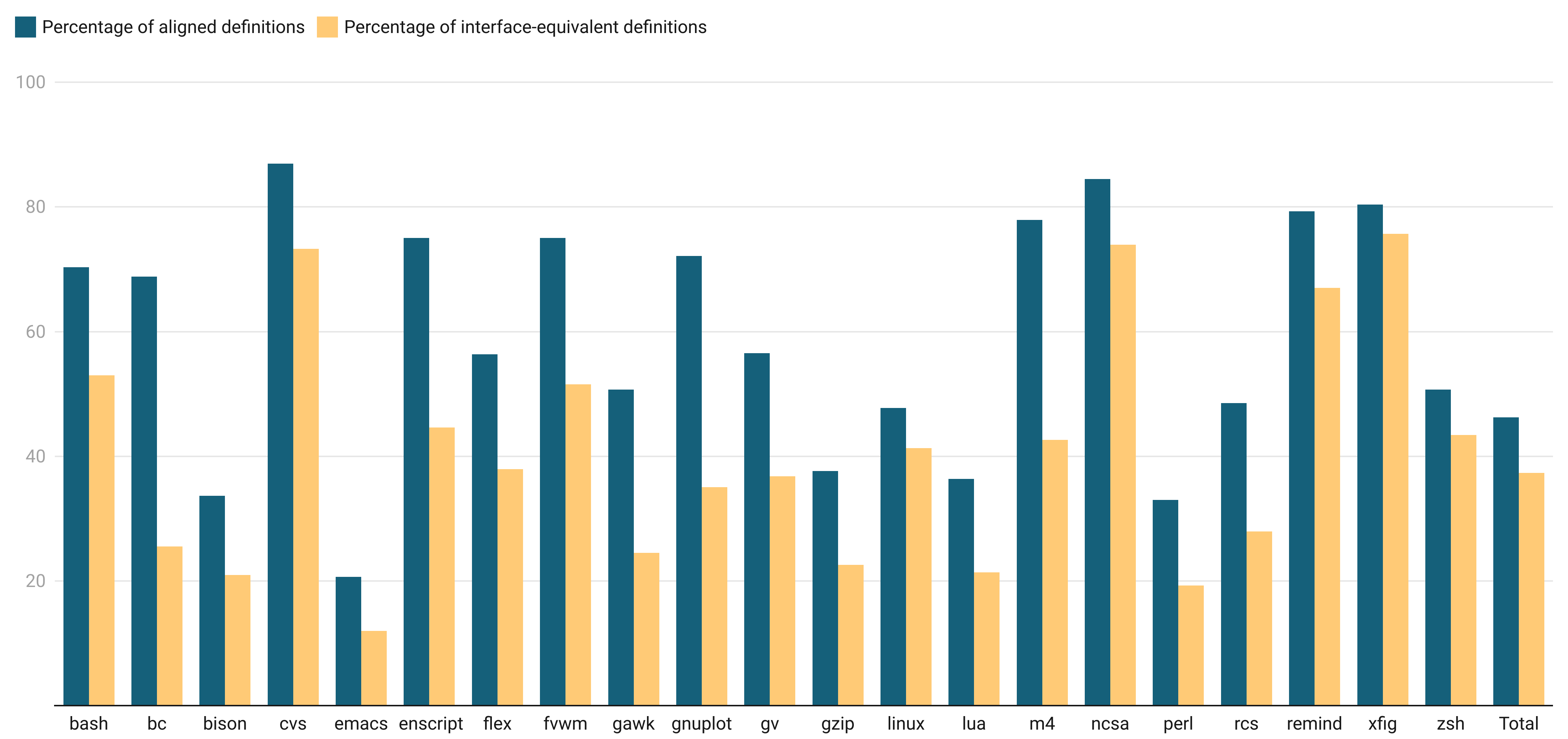}
\caption{Percentage of aligned and interface-equivalent definitions in each program.}
\label{fig:definition-stat-groups}
\end{figure*}

The C preprocessor is lexical and has no requirement to respect C syntax.  This research question evaluates the portability of C macro usage when only considering syntactic macro use to see the impact on portability.
To measure syntactic macro usage, we measure what C syntax the macro usage generates (statements, expressions, etc.) and whether it respects C syntax or violates it, which corresponds to the Unaligned macro property (Table~\ref{tab:properties}).

Figure~\ref{fig:definition-stat-groups} presents the percentage of each program's macro definitions that align with the program's AST, shown in blue, compared with its percentage of interface-equivalent macro definitions, shown in gold.
Across all programs, the percentage of aligned definitions necessarily exceeds the percentage of interface-equivalent definitions, because interface-equivalence depends on syntactic macro usage (the Unaligned property precludes interface-equivalence).

In general, the majority of syntactic macro definitions are easily-portable in most programs.
Linux, for instance, has a very large portion that are interface-equivalent, likely because of the good macro usage coding guidelines encouraged by maintainers~\cite{linuxcodingstyle}.
In some programs, however, only a minority of syntactic macros are easily-portable.
For instance, \gnuplot has many more aligned definitions than interface-equivalent ones because 8\% of its definitions align with types (more than any other program), and due to limitations with Clang we cannot fully analyze the semantic properties of type-aligned invocations and conservatively report them as not interface-equivalent.

\subsection{RQ4: Runtime}

We measure the end-to-end runtime to analyze all definitions' properties and portability categories for each program.
Across all \numevaluationprograms programs in our benchmark, the median time necessary to analyze all macro usage was about nine minutes and 20 seconds.
The analysis of Linux took the longest, requiring about 17 days and 21 hours, although it was run with more threads than the rest of the programs.
In total, it took about 19 days and eight hours to fully analyze all 86,199 macros across all programs.

\subsection{RQ5: Accuracy}
\label{sec:accuracy}
We measure Maki's accuracy by comparing it against ground truth for a statistically significant sample from the benchmark.
A 383-macro sample from the 86,199 macros in our benchmark has a 5\% margin of error and 95\% confidence.
We created ground truth for the sample by hand-checking all portability categories shown in Table~\ref{tab:properties} for each macro.
We compute the precision ($\frac{\text{true positives}}{\text{true positives}\, + \,\text{false positives}}$) and recall ($\frac{\text{true positives}}{\text{true positives}\, + \,\text{false negatives}}$), then we compute accuracy as the $\text{F}_1$ score (harmonic mean of precision and recall).
Maki has a true positive when it reports the same portability categories as ground truth, a false positive when the portability categories differ.
It has a true negative when the absence of portability categories matches ground truth, a false negative if it fails to report portability categories from ground truth.

Out of the 383-macro sample, Maki had 337 true positives, 17 false positives, and 26 false negatives, resulting in a precision of 95\%, a recall of 93\% and an $\text{F}_1$ accuracy of 94\%.

Seven false positives were due to the macro expanding to a designated initializer~\cite{gcc-designated-inits} or a statement expression~\cite{gcc-statement-exprs}, GCC-specific C language extensions that Maki does not yet support.
Maki appears to have aligned the ten remaining false positive macros with incorrect AST subtrees, leading to misclassifications.
Maki failed to match 23 of the false negatives with any portability category at all, and failed to identify the remaining three as having nested invocations.
By careful manual inspection we identified correct portability levels for all these macros, and therefore attribute these errors to limitations in Maki's implementation rather than the underlying framework.
We plan on adding these misclassifications to Maki's test suite so that we may resolve them.

A spreadsheet of the sample detailing ground truth creation can be found in our publicly-available artifact~\cite{artifact}.

\section{Case Studies}
\label{sec:case-study}
We present three case studies to demonstrate the capabilities of the Macro Inspector Framework and of Maki.
First, we run Maki on Linux's driver staging directory, which contains drivers under development, and hand-transform several macros Maki identifies as easy-to-port.  We submit these changes as patches to the Linux Kernel Mailing List for approval~\cite{lkml} and report on the discussion that led to their acceptance or rejection by developers.
This case study shows the viability of using Maki to help development on an actively-maintained, large-scale, critical C codebase that has a rigorous review process.

Second, using Maki's output as a guide, we transform by hand all macros in two programs from our benchmarks, enscript and m4, as well as two Linux modules, \texttt{ipc} and \texttt{sound/atmel}.
This case study shows the utility of Maki to help quickly port away from macro usage.

\subsection{Patching Linux Kernel Macro Usage}
\label{sec:linux-case-study}

Modifying the Linux kernel requires negotiating a careful and sometimes lengthy patch review process with the Linux developers~\cite{linuxpatches}.  We therefore target a select set of macros in the \texttt{driver/staging/} directory~\cite{staging} as candidates for patching in the Linux kernel.
We selected the first 11 macros that Maki identifies as definition-adapting;
definition-adapting are the simplest to port, requiring only a change of syntax from a macro definition to a C function with no further refactoring.
For example, \verb'GDM_TTY_READY' from \texttt{gdm724x/gdm\_tty.c} evaluates a conditional expression for if-statements:
\begin{lstlisting}[style=CStyle]
#define |\highlightcode{GDM\_TTY\_READY(gdm)}| \
  (gdm && gdm->tty_dev && gdm->port.count)
struct gdm *gdm = tty_dev->gdm[index];
if (!|\highlightcode{GDM\_TTY\_READY(gdm)}|);
\end{lstlisting}
Porting this macro requires only copying the macro body to the new function's return statement.  Since Maki confirms definition-equivalence, this transformation is safe for all invocations:
\begin{lstlisting}[style=CStyle]
static inline bool gdm_tty_ready(struct gdm *gdm) {
	return |\highlightcode{gdm \&\& gdm->tty\_dev \&\& gdm->port.count}|;
}
\end{lstlisting}

In total, we submitted eight patches transforming 11 macros.
As of writing, Linux maintainers have accepted six of our patches which port nine macros.  The remaining two macro transformations were rejected, because the maintainer preferred keeping the macro.  In one case, the maintainer preferred permanently inlining it, making porting unnecessary.

Discussion with the maintainer on two of the ported macros led to safer code:
First, \texttt{FPNTBL\_BYTES} from \texttt{media/atomisp/pci/sh\_css\_params.c} contained unsafe multiplication susceptible to overflow.  The maintainer requested using the \verb'array3_size' helper, which computes size without overflowing~\cite{memalloc}.
Second, \texttt{irq\_data\_to\_gpio\_chip()} from \texttt{greybus/gpio.c} expanded to a \texttt{void *} expression.
But the maintainer observed the macro's return value is only ever assigned to variables of type \verb'struct gpio_chip *' and requested the new function have this return type.
All patches we submitted, along with their mailing list histories, are included in the publicly-available artifact~\cite{artifact}.

\subsection{Porting All Macros in a Codebase}
\label{sec:porting-case-study}

We hand-ported macros to C functions in two codebase from our benchmarks, m4 and enscript.
For each codebase, we used Maki's output to produce a table of all macros it encountered, sorted by the portability categories listed in Table~\ref{tab:properties}.
We transformed each of these macros to C, one-at-a-time, starting with the interface-equivalent, which are the easiest to port.
After each transformation, we ensured the program still built (`make`) and passed tests (`make check`).

As shown in Figure~\ref{fig:definition-transformation-types}, m4 has 68 macro definitions and enscript has 56.
Roughly half of macro definitions are of interface-equivalent macros in both codebases, 29 for m4 and 25 for enscript.
These macros were fast and easy to transform by hand, taking one author only 20 minutes for enscript and 15 minutes for m4 to transform all the interface-equivalent macros.
Maki enables this speed, because it rules out complex macro behavior for these macros, allowing for a simple syntactic conversion. 

The non-interface-equivalent macros were more time-consuming.
These macros (39 in m4 and 31 in enscript) took about 3:40 hours and 4 hours total to transform by hand, respectively.
The challenge in porting is determining how to preserve the macro's non-functional behavior in C.
For instance, the macro \texttt{\_\_P}, defined in enscript, is conditionally defined to either expand to a given parameter list or an empty parameter list.
\begin{lstlisting}[style=CStyle]
#if PROTOTYPES
#define ___P(protos) protos
#else /* no PROTOTYPES */
#define ___P(protos) ()
extern char *strerror ___P ((int));
\\ 49 more function declarations.
\end{lstlisting}
It is used to remove the parameter list for older versions of C.  We converted each unique function signature into a conditionally-declared typedef to include or exclude parameters.

We found that Maki's results had two false negatives in enscript and 12 in m4.
The two in enscript and eight in m4 were due to Maki's lack of support for ad-hoc polymorphism and variadic arguments in macro invocations.
The remaining four Maki identified as calling-convention-adapting because they modified a bit-field argument.
The C standard prohibits taking the addresses of bit-fields~\cite{ISO-IEC-9899-2011}, a subtlety of C that Maki does not take into account.

We also hand-ported two whole Linux modules, \texttt{ipc} and \texttt{sound/atmel}, each containing 49 and 55 macros, respectively, taking three hours total.
91 macros were interface-equivalent while the remaining 13 were non-interface-equivalent.
We transformed 102 out of the 104 macros defined in both modules; the two macros we did not transform were header guard macros.
We checked the changes by compiling the modules and running Linux's checker, \texttt{sparse}, with \texttt{make C=2}.
We have not submitted patches for these changes to Linux maintainers because they are significantly larger in scope than the patches discussed in Section~\ref{sec:linux-case-study}, and Linux maintainers prefer to review series of small patches rather than large patches~\cite{lkml-small-patches}.

The changes made to all transformed programs and detailed notes can be found in our publicly-available artifact~\cite{artifact}.

\section{Maki for Code Generation}
\label{sec:code-generation}
In addition to guidance for hand-porting, \tool can help inform mechanical transformation approaches by providing guarantees about macro usage.
When function-like macros are definition-adapting, the semantics of the macro are identical to a C function, requiring only simple syntactic changes to place the macro body in a return statement and to replace the \verb'#define' with a C function prototype, as shown in the Linux Case Study (Section~\ref{sec:linux-case-study}).
All other interface-equivalent macros are either calling-convention-adapting or scope-adapting, i.e., they differ from functions in calling-convention (call-by-name vs. call-by-value) or scoping.

When macros are calling-convention-adapting, i.e., they exploit macros' call-by-name semantics, generating C code versions of the macros only requires simulating call-by-name with call-by-value.
For instance, when there are side-effects on a macro's argument, such as \verb'PREPEND_LIST' (Section~\ref{sec:introduction}), the call-by-name convention means the updated value of the argument is reflected in the caller.  Such behavior can be simulated by passing references to the arguments instead of their values, e.g., passing a pointer in C, passing-by-reference in C++, or passing an object in Java.

Scope-adapting macros exploit the lack of any scope, besides global, and the dynamic scoping of macro definitions.  Macros' lack of scoping enables them to capture symbols from their callers' scope as seen in the use of the function-local \verb'sp' variable in the \verb'ISSET'  macro (Section~\ref{sec:framework}) without explicitly passing a parameter.
Maki's ability to identify scope-adapting properties would enable automated code generation tools to identify what refactorings are needed to produce equivalent functions.
For macros that exploit the lack of scoping, the code generation tool need only identify any undeclared symbols in the macro body that are in the caller's scope, then add these as parameters, akin to a function extraction refactoring~\cite{refactoring}.
Macros that exploit dynamic scoping only need reorderings of declarations by the code generation tool to ensure any macros called in the body are available in the static scope.

As the hand-ported macro case study shows (Section~\ref{sec:porting-case-study}), non-interface-equivalent macros are more complicated to port, because they are not guaranteed to correspond one-to-one with a function.  They instead require first refactoring the functional interface to match the language's function semantics before transforming the macro.
For instance, when macro arguments have side-effects, e.g., a macro takes \verb'x++' as an argument, this side-effect is repeated every time the argument is used.
A code generation tool can simulate this behavior by refactoring the argument into a thunk, which preserves the side-effect throughout the execution of the macro body and reflects the final value in the caller's scope.

Even though functions take expression parameters and function calls themselves are expressions, macros may take and return any grammar construct, including declarations and control-flow altering statements such as \verb'goto' or \verb'break'.
Barring language-specific extensions to functions such as C's \emph{longjmp}, a code generation tool will likely not be able to simulate these behaviors with functions in general.
For nested macros, however, there is more hope for code generation tools.
When the outer macro is interface-equivalent, it may be possible to first convert the outer macro definition, thereby unnesting the inner macro.  After porting, the inner macro is now an outer macro that can have its properties checked for portability.

\section{Threats to Validity}
\label{sec:validity}

\emph{Internal validity.}
The properties are intended cover all macro usage.
If any one macro is not analyzed, we might miss portability.
To ensure completeness, we designed the properties so that when a macro meets none of the properties, it is interface-equivalent (the definition-adapting portability category).
To ensure that the implementation of our macro analysis framework in \tool matches the formal properties, we developed a test suite to exercise each property, made from hand-crafted examples and tests adapted from benchmarks.  In total, our test suite comprises 63 files and 895 source lines of code.
We used this test suite to make Maki highly accurate, as Section~\ref{sec:accuracy} demonstrates.
We intend to add the misclassified macros we found while hand-checking Maki's output against ground truth in Section~\ref{sec:accuracy} to Maki's test suite so that we may address them as well.
Finally, due to a limitation with Clang, Maki is unable to check if type-aligned macro invocations satisfy any scope-adapting properties, so we conservatively assume all type-aligned invocations are scope-adapting.
Since scope-adapting macros are interface-equivalent, even if this issue were resolved \tool would still find the same quantity of interface-equivalent macros.

\emph{External validity.}
The results about macro usage and portability depend on the set of benchmark programs chosen.
To achieve a wide variety of program types, we started with Ernst et al.'s preprocessor metrics benchmarks, which has programs from many domains, including languages, utilities, shells, etc.
To broaden the range and size of programs, and to compensate for older programs no longer under development, we added new benchmarks.
Our framework is geared towards C preprocessor macros, but other macro systems, such as for Rust~\cite{rustmacros} and Lisp~\cite{lispmacros} have different semantics.
Some properties from our framework would apply, such as scoping differences, but applying our framework to other macro systems would require adjusting and adding properties, which we leave as future work.
Moreover, porting tools that target high-level language constructs may have one-to-one mappings for non-interface equivalent macros; for instance, languages with first-class functions could map thunkizing macros using anonymous functions.
Similarly, we do not support analyzing macros used in C++ programs, which could have different properties, e.g., related to its additional language constructs that would affect how often macros are easily-portable.
On the other hand, porting to object-oriented languages could also open the door to more types of macro transformations, which we leave to future work.

\section{Related Work}
\label{sec:related-work}

The most recent and only related macro property analyzer is Mennie and Clarke's~\cite{mennie} automated tool for transforming certain object-like macros to C variables.
Their tool collects facts about macros, classifies them based on these facts, and then generates plans for transforming each macro based on its facts and classifications.
It employs sophisticated rules for inserting transformed code, which enable it to automatically transform certain object-like macros into correctly-scoped local variables.
\tool's design is similar to theirs in that it first collects properties about macros, and then categorizes macros based on the properties they satisfy.
\tool differs from Mennie and Clarke's work in that it does not automatically transform macros, and analyzes a wider array of macro properties to find many more easily portable macros.

SugarC~\cite{sugarc} transforms preprocessor usage to C by targeting the preprocessor static conditionals and convering them into runtime C conditions. %
SugarC improves variability-aware analyses of programs that use preprocessor static conditionals for configuration management, but does not try to maintain developer abstractions in the transformed code it produces.
Hercules~\cite{hercules} is another tool that transforms CPP compile time conditionals to C runtime conditionals.
Unlike SugarC, it uses an AST generated by Typechef~\cite{typechef} to perform its transformation.
C Reconfigurator's~\cite{creconfigurator} transformation rules were proven to be sound, but only for a theoretical language that is a subset of what it actually transforms.

McCloskey and Brewer~\cite{astec} developed Macroscope to transform CPP macros to new a macro preprocessor language, called ASTEC.
ASTEC has advantages over CPP, but presents all the same issues to porting tools as CPP code since it is still a preprocessor language.
Moreover, while Macroscope can merge duplicate transformed definitions into single definitions, their rules for placing transformed definitions are not as rich as those of Mennie and Clarke's tool~\cite{mennie}.

The C preprocessor analysis tool most relevant to this project is Dietrich's CppSig~\cite{dietrich}, which collects macro invocations into tree structures, and infers function signatures for them.
\tool relies on insights akin to those behind CppSig to find AST-aligned macro invocations and infer types for invocations' expansions and arguments.
The seminal work in CPP analysis is that of Badros and Notkin~\cite{cppanalysis}, which outlines a method of C source code analysis that offers both preprocessing and C parsing ``actions'' that are analogous to Clang's preprocessor callbacks~\cite{clang-ppcallbacks} and AST~\cite{clang-ast}.
This is similar in spirit to the approach taken by \tool, since it uses both Clang preprocessor callbacks and the Clang-generated AST to perform its analysis.
The first large study of macro usage was conducted by Ernst et al.~\cite{ernst}, from which we drew many of our benchmarks programs.
CScout~\cite{cscout} enables the analysis of ``program families''~\cite{cscout}, i.e., workspaces comprised of interdependent programming languages.
CScout can perform simple refactorings on C program families such as identifier renaming, but cannot identify all the easily portable macros that \tool does.

Favre's~\cite{favre} is the only work we know of to fully formalize the preprocessor itself.
He outlined a denotational semantics for CPP, taking into account subtle features such as  stringification and tokenization~\cite{cppmanual}. %
Since Favre does not consider macro's interaction with C code, it is unrelated to the AST-oriented properties of our the Macro Inspector Framework.

There have been several proposals for hygienic alternatives to preprocessor macros for C~\cite{ace, weise, astec}, including the aforementioned ASTEC~\cite{astec}.
These macro languages are syntactic, and require that all macro invocations align with the program's AST; however they are still preprocessor languages that similarly hinder porting.
Rust~\cite{rust} offers syntactic macro where arguments are annotated with their AST-node type.
Syntactic macro systems have a home in the Lisp family of programming languages~\cite{racket-lang, flatt}.
Pombrio and Krishnamurthi~\cite{resugaring} demonstrated the feasibility of reconstructing abstractions in Lisp-like languages.

3C~\cite{3c} ports C programs to Checked C~\cite{checkedc}, a pointer-safe dialect of C designed to prevent memory bugs.
Macro usage has impaired 3C porting efforts~\cite{3cissue40,3cissue439,3cissue400}.
\tool could be used to help develop porting tools to mitigate the challenge.
c2rust~\cite{c2rust} translates C code to unsafe Rust code.
Like 3C, c2rust faces problems with preprocessor macros~\cite{c2rustissue16} and preprocesses first, losing macro abstractions.

\section{Conclusion}
\label{sec:conclusion}
In this work we present the Macro Inspector Framework and its embodiment in the Maki analyzer.
Compared to prior work, our implementation finds an average of twice as many easily portable macros, and up to 7x more for programs with more complex macro usage.
Using our framework as a guide, we hand-patched 11 \linux kernel macros, nine of which have been accepted by kernel maintainers.
In future work, we will study the application of our framework's properties when used to port macros to other languages besides C.
Ultimately, we plan to leverage our framework to create an automated tool for porting easy-to-port macros, so that developers can focus their porting efforts on more complex definitions.

\section*{Acknowledgments}
\label{acknowledgment}
We would like to thank Mike Hicks, Elaine Weyuker, and all the reviewers for their valuable feedback.
This work was supported in part by NSF grants CCF-1840934 %
and CCF-1941816. %

\bibliographystyle{acm}
\bibliography{references.bib}

\end{document}